\documentclass{PoS}

\title{H.E.S.S. highlights}

\ShortTitle{H.E.S.S. highlights}

\author{\speaker{Gerd P{\"u}hlhofer}%
\\
       Institut f{\"u}r Astronomie und Astrophysik, Eberhard Karls Universit{\"a}t T{\"u}bingen \\
       E-mail: \email{gerd.puehlhofer@astro.uni-tuebingen.de}}

\author{for the H.E.S.S. collaboration\\
        E-mail: \email{contact.hess@hess-experiment.eu}}

\abstract{The H.E.S.S.\ collaboration continues to run an array of five Imaging Atmospheric Cherenkov Telescopes to observe the Southern sky in very high energy gamma-rays. In this presentation, recent highlight results obtained with H.E.S.S.\ are briefly reviewed, with references to the relevant publications.}

\FullConference{7th International Fermi Symposium\\
         15-20 October 2017\\
         Garmisch-Partenkirchen, Germany}

\begin{document}

\section{Introduction: The H.E.S.S.\ experiment}

\subsection{H.E.S.S.\,phase I and II}
The High Energy Stereoscopic System (H.E.S.S.) is an array of Imaging Atmospheric Cheren\-kov Telescopes located in the Khomas Highland of Namibia at an altitude of 1800\,m above sea level. Four 12\,m telescopes (CT1-CT4) are in regular operation since 2003 in stereoscopic operation (H.E.S.S.\ phase I). Their cameras have a field of view (FoV) of $5^\circ$ in diameter, which translates into a celestial FoV for reconstructed $\gamma$-rays of $3^\circ$ diameter (with $\gtrsim 70\%$ of peak acceptance). The four-telescope array has an energy threshold of $\sim$$100$\,GeV at zenith. In 2012, a fifth telescope (CT5) was added to the array, marking the beginning of H.E.S.S.\ phase II. The telescope has a $\sim$$28$\,m diameter mirror, the camera a field of view of $3.5^\circ$. With the new telescope, the energy threshold of the array is substantially lowered (with the analysis threshold value being adapted to the signal-to-noise ratio of the respective data set), but with the restriction to a smaller FoV at low energies. 

The five telescopes run with a hybrid-array trigger. To exploit the low energy threshold of CT5, the 28\,m telescope triggers event readout independently of the participation of the 12\,m telescopes. In regular array operation, the majority of all events are such ``mono'' events. With the original CT1-CT4 electronics (i.e.\ before the upgrade, see Sect.\,\ref{SectHESS1U}), $\sim$$20\%$ of all events have been triggered simultaneously by CT5 and (at least) one of the smaller CT1-CT4 telescopes. Only a small fraction of events are triggered by only (at least two of) the 12\,m telescopes. Mono events suffer from a reduced background rejection fraction compared to stereo events, and $\gamma$-rays can only be reconstructed with a poorer angular (and energy) resolution. Together with the higher participation rate this leads to the fact that the corresponding low-energy range of H.E.S.S.\ II (which provides the best energy overlap with {\it Fermi-}LAT) is systematics-limited in data sets with longer exposures. Therefore, depending on the source flux and spectrum, increasing the observation time of steady sources beyond a certain level does not improve the quality of the reconstructed source parameters. On the other hand, due to the huge effective area of the instrument, temporal structures of reasonably bright sources can be resolved with good photon statistics. In the higher energy regime where stereo events start to dominate, observations are for typical exposure times statistics-limited like in the H.E.S.S.\ I phase.

\subsection{H.E.S.S.\ CT1-CT4 camera upgrade}
\label{SectHESS1U}
Besides the installation of CT5, a complete upgrade of the camera electronics of CT1-CT4 was the most significant servicing that was performed in the past years for the continued successful operation of the H.E.S.S.\ experiment. The upgrade employed new electronics concepts that were developed in view of CTA. The main goal of the upgrade was to replace the existing and aging electronics with a lower deadtime readout system. This permits to run the hybrid-array trigger with a significantly lower energy threshold for the events in which CT5 and one of the CT1-CT4 cameras participate (the corresponding trigger fraction has increased to above $40\%$), thus permitting a broader energy overlap between mono (CT5 only) and stereo ($\ge 2$ telescopes from CT1-CT5) events. The new cameras were installed in 2015-2016, commissioning ended early 2017~\cite{bib:hess1ucameras2017I,bib:hess1ucameras2017II}.

\section{Recent selected H.E.S.S.\ II highlights}

\subsection{Active Galactic Nuclei}

Very high energy (VHE) AGN spectra are strongly affected by attenuation by the extragalactic background light (EBL). With increasing redshift, spectra become on average softer and ultimately undetectable above a given detector energy threshold. Lowering the energy threshold of H.E.S.S.\ with the inclusion of the 28\,m telescope therefore substantially expands the visible universe for H.E.S.S.\ observations. First results on the well-known VHE-emitting blazars PKS\,2155-304 and PG\,1553+133 using H.E.S.S.\,II mono data have been published in~\cite{bib:hess2017agnmono}. The energy spectra show a smooth continuation to appropriately temporally-sampled {\it Fermi-}LAT spectra at lower energies, once the VHE spectra are corrected for EBL absorption using an up-to-date EBL model in the relevant wavelength range.

\subsection{The Vela pulsar}

Another obvious source class to investigate, also with the goal to verify the performance of H.E.S.S.\ II at low energies, are pulsars that are known to emit high energy pulsed emission as seen with {\it Fermi-}LAT (or EGRET). After the detection of VHE $\gamma$-rays from the Crab pulsar, an obvious target for H.E.S.S.\ was therefore the Vela pulsar. The phasogram obtained with H.E.S.S.\ II mono data shows a clear detection of photons in the phase bin called ``P2'', with $\sim$16,000 photons above background. The background level in the H.E.S.S.\ phasogram is however substantial, the signal-to-noise level in P2 corresponds to a detection significance of $\gtrsim 15\,\sigma$. The light curve in P2 is well compatible with the corresponding light curve measured with {\it Fermi-}LAT above 10\,GeV. Of high interest is the spectral shape of the VHE emission up to the highest detectable energies, and potentially a change of the pulse profile with energy. More details will be published in a forthcoming journal article.

\subsection{The colliding wind binary system Eta Carinae}

Another obvious target of interest for H.E.S.S.\ II is the colliding wind binary system Eta Car. Previous H.E.S.S.\ I observations have only yielded upper limits. {\it Fermi-}LAT observations have revealed that Eta Car is a high-energy $\gamma$-ray emitter. The system has been observed with H.E.S.S.\,II pre-periastron and around periastron in summer 2014. The observations and the data analysis are particularly challenging because of the high optical photon noise from stars in the field around Eta Car. After a careful analysis, the H.E.S.S.\,II data set showed a clear signal (total significance of $>13\,\sigma$)~\cite{bib:hess2017etacar}, the spectral analysis reveals a soft VHE spectrum with a smooth connection to the {\it Fermi-}LAT spectrum below 100\,GeV.

\section{H.E.S.S.\ II and the transient VHE sky}

Because of the technical reasons outlined above, one main topic when exploiting the low energy threshold of H.E.S.S.\ CT5 is the study of time-variable sources. Physics examples are the observations of pulsars, of binary systems, and AGN monitoring. Many events are of non-predictable nature, therefore corresponding H.E.S.S.\ observations may need to be triggered by other instruments, mostly having strong survey capabilities. One prominent example is the blazar target of opportunity (ToO) program~\cite{bib:hess2017blazartoo}. Further programs are sketched in the following.

\subsection{H.E.S.S.\ observations of gamma-ray bursts}

Since early in the H.E.S.S.\ I phase, the H.E.S.S.\ collaboration has conducted an elaborate gamma-ray burst (GRB) follow-up program. With H.E.S.S.\ II, the response time of H.E.S.S.\ to such alerts has been significantly reduced. Besides improvements in the trigger processing and data acquisition scheme for all telescopes, CT5 was specifically designed for improved response times. Due to the fact that CT5 can go in reverse tracking mode for burst follow-ups (i.e.\ the telescope can be driven with the altitude axis through zenith), $>90\%$ of the targets can be reached by the telescope within 60 seconds. Besides observations of the prompt phase of GRBs (in VHE observational terms defined as observations within minutes after the initial burst), also afterglow observations are performed (on scales of hours to a day after the burst). Despite all efforts, GRBs have however escaped detection in the VHE band so far~\cite{bib:hess2017grbs}.

\subsection{Fast radio bursts}

Another source class of increasing interest are fast radio bursts (FRBs). Since a few years, a follow-up program is conducted with H.E.S.S.\ to observe FRBs which are detected, for example, with the Parkes telescope. Of particular interest was the event FRB\,150418, because of the potential identification of a host galaxy (with known redshift $z=0.492$). H.E.S.S.\ observations started 15 hours after the event, the derived upper limits constrain the non-thermal emission of the afterglow of this FRB~\cite{bib:hess2017frb150418}.

\subsection{Neutrino events}

Besides ToO observations of events triggered by observations in the electromagnetic (EM) spectrum, other messengers play an important role in the H.E.S.S.\ ToO program. High-energy neutrino events provided by the ANTARES and IceCube experiments are followed up~\cite{bib:hess2017multimessenger,bib:hess2017icecube}, with an increasingly automated procedure\footnote{see e.g.\ https://indico.desy.de/contributionDisplay.py?contribId=35\&confId=14253}. An exciting event was recently followed up with H.E.S.S.\ as well as with other Cherenkov telescopes: IceCube-EHE-170922. The H.E.S.S.\ collaboration issued an ATEL on the 27$^\mathrm{th}$ of September 2017, reporting upper limits derived across the positional error box of the neutrino event. However, contemporaneous $\gamma$-ray activity of the AGN TXS\,0506+056 reported by the {\it Fermi-}LAT and MAGIC collaborations in the high energy (HE) and VHE band, respectively, was used to identify the AGN as the potential source of the high-energy neutrino. Therefore, observations with H.E.S.S.\ also continued. Results will be reported in a forthcoming journal paper.

\subsection{Gravitational wave events}

\begin{figure}[t]
    \centering
	\includegraphics[width=0.9\textwidth]{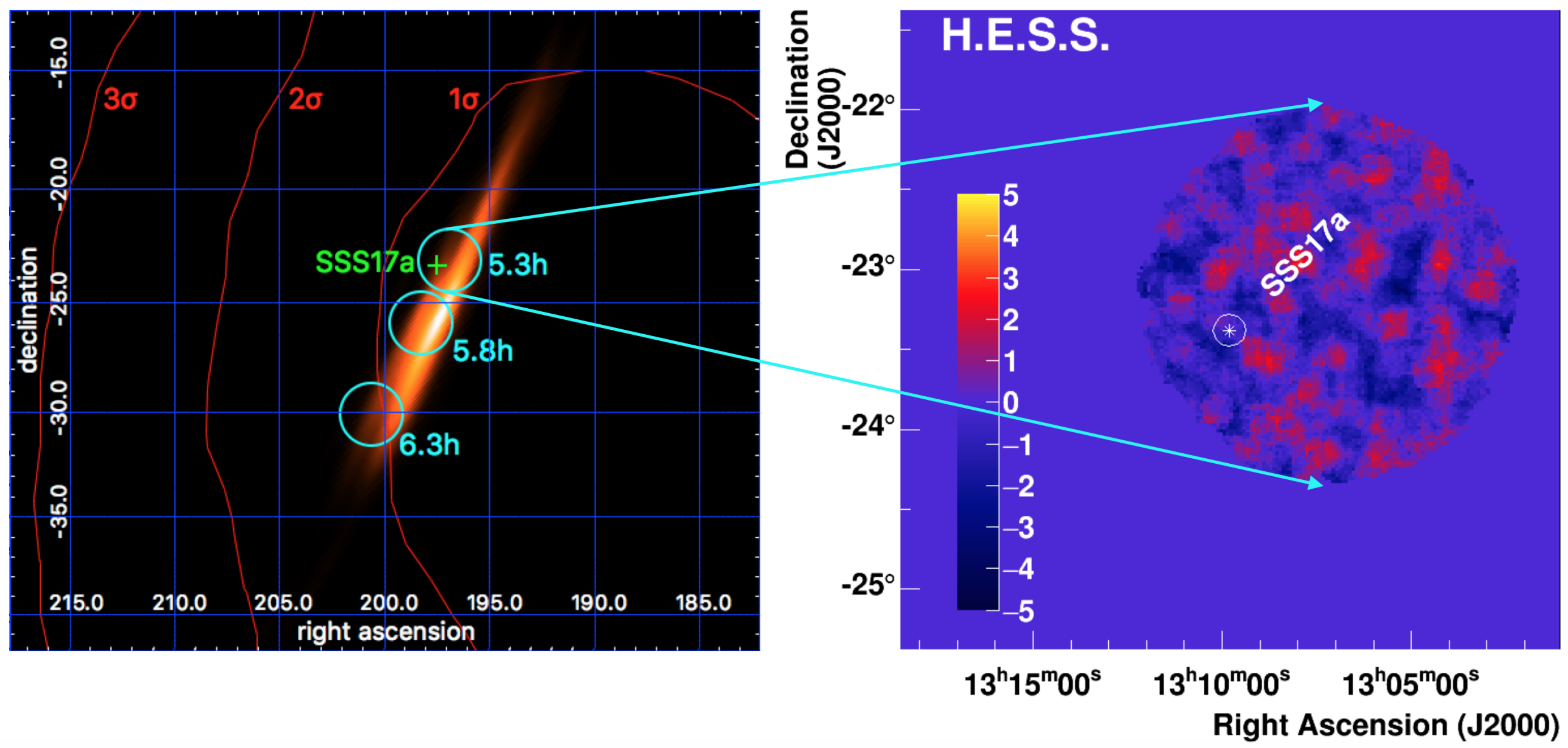}
	\caption{
Left: Pointing directions of the first night of H.E.S.S.\ follow-up observations of GW\,170817. The circles denote the H.E.S.S.\ FoV (using a diameter of $3^\circ$). Shown are the starting times with respect to GW\,170817. The colored map is a representation of the directional reconstruction of the GW event (the LALInference map provided by the GW pipelines). Red lines are part of the localisation of GRB170817A. Right: Significance map of the first H.E.S.S.\ pointing covering SSS17a. The white circle (with a diameter of $0.1^\circ$) denotes the H.E.S.S.\ psf and the size used for the oversampling of the map.
Figure taken from~\cite{bib:hess2017GW170817}.}
	\label{Fig:HESSGWSkyMap}
\end{figure}

The most exciting recent event triggered by a non-EM detector was of course the gravitational wave (GW) event GW\,170817, detected in coincidence with the GW detectors LIGO-Livingston and LIGO-Hanford and simultaneously observed with Virgo, which just recently had been included in the joint LIGO-Virgo observational run. For the first time, a neutron star (NS)-NS merger event was detected using gravitational waves, and at the same time the positional uncertainty of the reconstructed direction from where the gravitational waves were emitted was greatly reduced due to the additional information from Virgo. Besides the detection of a short GRB (GRB\,170817A) as temporal (and positional) counterpart just $\sim$$2$ seconds after the event with {\it Fermi-}GBM and the INTEGRAL anticoincidence detector, the localisation of the GW event permitted to identify also a delayed EM counterpart in the optical (SSS17a, identified as a kilonova) as well as in the radio and X-ray bands. The H.E.S.S.\ experiment participates in the world-wide EM-follow up program initiated by the LIGO-Virgo-consortium. H.E.S.S.\ observations started 5.3 hours after GW\,170817, using the GW error box as well as a catalog of nearby galaxies to define an observation plan (see Fig.\,\ref{Fig:HESSGWSkyMap}). Due to this strategy, the later-identified counterpart was indeed covered in the FoV of the first H.E.S.S.\ pointing. In retrospect, H.E.S.S.\ was the first ground-based telescope following up GW\,170817. Upper limits were obtained~\cite{bib:hess2017GW170817} and added to the joint description of the EM follow-up observations performed on GW\,170817~\cite{bib:mwl2017GW170817}.

\section{Probing cosmic-ray accelerators on different scales}

One of the primary goals of (HE and VHE) $\gamma$-ray observations of celestial objects is to identify and study cosmic ray (CR) acceleration in these objects. Gamma-ray detectors (satellite-based like EGRET, AGILE, and {\it Fermi-}LAT) as well as ground-based (Cherenkov telescopes like H.E.S.S., MAGIC, and VERITAS, as well as water Cherenkov detectors like HAWC) have detected in the past few decades a large number of celestial $\gamma$-ray sources. For the study of CR acceleration, Galactic sources are of prime interest, since the study of CR acceleration up to energies of the knee in the CR spectrum (up to which the spectrum is supposed to be dominated by Galactic CRs) can best be studied in these sources. However, while $\gamma$-rays probe parent particles of similar energy, the photons don't carry direct information whether they were produced in hadronic processes (i.e.\ they are tracing primary hadrons) or in leptonic processes (i.e.\ they are tracing primary leptons). This ambiguity is of high importance (and often a hindrance) when studying the efficiency and the energy budget of the particle acceleration.

With VHE observations, the acceleration to the highest energies directly accessible with $\gamma$-rays can be studied, as well as the injection of CR particles at the upper end of the spectrum from the sources into the surrounding ISM. The ensemble of CRs in our Galaxy (after diffusing away from the sources) can best be studied with large FoV detectors at HE, i.e.\ presently with {\it Fermi-}LAT. Also, the HE band provides a potential signature for the hadronic nature of a spectrum, namely the ``pion bump'' detected in the spectra of some sources. However, the probed energies are far below the knee energies. Therefore, the search for ``PeVatrons'' has emerged as a key tool to search for hadronic accelerators up to knee energies. Unbroken VHE spectra up to $\sim$$100$\,TeV would provide a clear signature for PeV (i.e.\ up to knee energies) acceleration and at the same time provide a signature for hadronic acceleration. However, even objects which do substantially contribute to the acceleration of Galactic CRs will most likely only show up as Pevatrons for a short amount of time, since confinement times for PeV particles are likely to be relatively short.

\subsection{The H.E.S.S.\ I legacy Galactic plane survey}

\begin{figure}[t]
    \centering
	\includegraphics[width=0.9\textwidth]{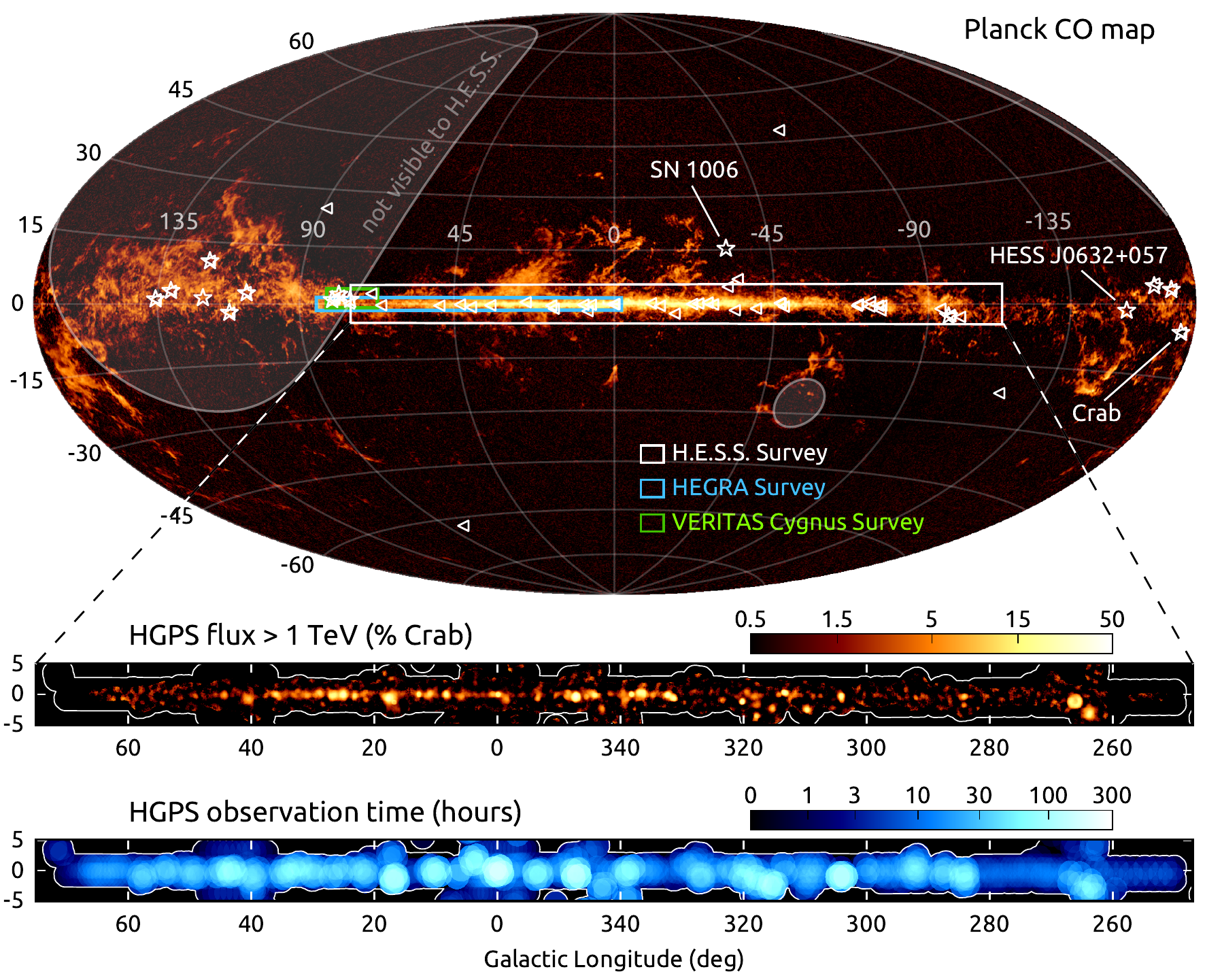}
	\caption{
An illustration of the H.E.S.S.\ I Galactic plane survey. The top panel shows the HGPS region on top of the carbon monoxide gas sky map measured with the Planck satellite. Triangles denote Galactic $\gamma$-ray sources measured at lower but complementary energies, recently detected with the {\it Fermi}-LAT satellite. Stars indicate the 15 Galactic VHE sources outside the HGPS region. The regions covered by earlier surveys are also indicated. The middle and bottom panels show the measured VHE $\gamma$-ray flux and the observation time, respectively. The white contours mark the edge of the survey region. Figure taken from~\cite{bib:hess2017hgps}.
}
	\label{Fig:HESSHGPS}
\end{figure}

With the H.E.S.S.\ phase I array, the central part of the Galactic plane has been surveyed over several years to search for new, previously unknown VHE sources. Together with reobservations of new sources and fields of interest and with dedicated observations of known Galactic objects, the H.E.S.S.\ I data set between $110^\circ < l < 65^\circ$ (Galactic longitude) and $-3.5^\circ < b < 3.5^\circ$ (Galactic latitude) forms the H.E.S.S.\ Galactic plane survey (HGPS) data set (see Fig.\,\ref{Fig:HESSHGPS}). The data comprise 2673 hours of (good quality) observations, performed in the years 2004 to 2013. Several new source detections and studies of individual sources and Galactic plane source populations have been published over the past years. The HGPS catalog will comprise three firmly identified binary systems, eight supernova remnants (SNRs), 12 firmly identified pulsar wind nebulae (PWNe), and eight composite systems (i.e.\ the identified counterpart is a composite (SNR and PWN) system). 11 VHE sources have no known counterpart in other wavelength, while the largest group of sources (36) have associated counterparts but a firm identification is not possible~\cite{bib:hess2017hgps}.

\subsection{New VHE supernova remnants / remnant candidates}

The vast majority of the HGPS sources are extended beyond the point spread function of the data set (which is typically $R_{68\%} \simeq 0.07^\circ$). However, the morphological information from the VHE data alone is usually not sufficient to give strong clues to the nature of the source. The majority of unidentified sources are most likely PWNe, with often unspecific morphology. The HGPS data set permitted nevertheless to identify three new SNR candidates, using a shell-type appearance of the sources~\cite{bib:hess2017shells}. One of the sources (HESS\,J1534$-$571) was later confirmed as SNR, from the detection of a radio synchrotron counterpart. Opposite to that, HESS\,J1912+101 remains for the moment the only (likely) VHE SNR without counterpart. HESS\,J1614$-$465 has a well-known and clearly associated counterpart detected with {\it Fermi-}LAT, the information however does not permit to confirm the SNR classification.

Deeper observations of RCW\,86 also permitted to reveal the VHE shell-type morphology, further confirming the association of the VHE source with the well-known SNR~\cite{bib:hess2017rcw86}.

\subsection{Supernova remnants with a pion bump in their MeV-GeV spectra}

It is tempting to assume a hadronic nature of the VHE emission of the new SNRs mentioned above, at least for HESS\,J1534$-$571 for which deep {\it Suzaku} X-ray observations have not revealed any X-ray synchrotron emission. However, it is very challenging (and not achieved with the current data) to reach the necessary level of X-ray sensitivity for these large ($\sim$$0.4^\circ$) sources to be able to exclude a leptonic, inverse Compton (IC)-interpretation of the VHE emission. A morphological and spectral identification of the VHE source with a {\it Fermi-}LAT counterpart that exhibits a pion-bump signature for hadronic emission would provide clean evidence for hadronic emission up to the VHE regime. Indeed, this was successfully shown to be the case for the well-known SNR W49B~\cite{bib:hess2017w49b}. W49B is a rather young (1-4\,kyr) SNR which is interacting with dense molecular clouds ($n \sim 10^3$\,cm$^{-3}$). Connecting the supra-TeV protons to the GeV protons also helps for the discussion whether the GeV protons are freshly accelerated in the expanding SNR shock waves in these sources.

\subsection{RX\,J1713.7-3946: Escaping Cosmic rays?}

A thorough analysis of the VHE morphology of the well-known non-thermal SNR RX\,J1713.7-3946 and specifically a comparison with the non-thermal X-ray emission profile (measured with XMM-{\it Newton}) has offered another potential channel to discriminate hadronic from leptonic emission. VHE emission extending beyond the outer shocks as defined by the X-ray emission could in principle be taken as a signature of hadronic CRs escaping upstream of the shock. However, the interpretation is not unambiguous. E.g.\ a rapid drop of the magnetic field could also suppress the X-ray emission beyond a certain boundary while electrons could still persist and produce VHE IC emission. A detailed discussion can be found in~\cite{bib:hess2017rxj1713}.

\subsection{The search for PeVatrons}

So far, no VHE-emitting supernova remnant shows evidence for an unbroken VHE spectrum up to $\sim$$100$\,TeV, or evidence for a high-energy tail beyond a potentially leptonically-dominated spectrum at and below $\sim$$10$\,TeV. SNRs detected with sufficiently large VHE statistics exhibit cutoffs towards higher energies, most prominently seen at RX\,J1713.7-3946. A possible interpretation is that the confinement time of PeV particles in SNRs is so short that it is difficult to detect PeV particles in the established historical Galactic SNRs.

\subsubsection{Young supernovae}

In that context it is therefore interesting to explore the potential VHE emission from very young SNRs. Some theories predict that the particle acceleration to TeV energies and beyond may happen on very short time scales compared to the usual benchmark time scale, namely the transition to Sedov phase. Several positions of SN events within a year after their explosion have serendipitously been observed in the FoVs of AGN observations with H.E.S.S. No VHE emission has been detected from these sky locations, more details of that analysis can be found in~\cite{bib:hess2017youngsne}.

\subsubsection{Search for PeVatrons in the H.E.S.S.\ Galactic plane survey field}

Searches for Galactic PeVatron candidates are also performed amongst all HGPS catalog sources. Already a few years back, HESS\,J1641$-$463 has attracted attention for its hard spectrum (VHE photon index $\sim$$2.1$), and a derived lower limit of a potential cutoff in the spectrum of $\sim$$100$\,TeV, and has therefore been considered as a potential PeVatron. Recently, two further hard-spectrum H.E.S.S.\ sources, HESS\,J1741$-$302~\cite{bib:hess20171741} and HESS\,J1826$-$130, have been under thorough study for their potential PeVatron nature. At this point in time, however, alternative (non-hadronic) interpretations of the spectra of all three sources cannot be firmly excluded~\cite{bib:hess2017pevcantidates}.

\subsection{The Galactic center}

While the search for PeVatrons amongst SNRs and other H.E.S.S.\ survey sources has so far remained unsuccessful or inconclusive, the VHE source located at the Galactic center has revealed itself as a PeVatron through the injection of relativistic particles into the surrounding medium. While the VHE spectrum of the central point-like source (which is spatially coincident with the central Galactic black hole) cuts off at $\sim$$10$\,TeV, the surrounding diffuse emission shows no cut-off well above 10\,TeV and is consistent with a parent proton population up to 1\,PeV. The diffuse emission stems from the galactic ridge and can be shown to be induced by relativistic protons that have been injected by the central source and are diffusing outwards. A deeper study of the H.E.S.S.\ I data has meanwhile shown that the VHE emission can very precisely be decomposed into the expected different components in the central molecular zone (CMZ) and that the escaping CRs indeed fill the entire CMZ~\cite{bib:hess2017cmz}.

The Galactic center region has meanwhile also been deeply observed with the H.E.S.S.\ II array. The VHE spectrum of the central source (using stereo data only) connects well to the {\it Fermi-}LAT spectrum. However, the energy threshold is not low enough to cover the spectral break that is implied by connecting the {\it Fermi-}LAT and the H.E.S.S.\ spectra.

\subsection{NGC\,253: a calorimetric measurement of cosmic rays in a starburst galaxy}

The investigation of the ensemble of Galactic CRs is the domain of HE instruments like {\it Fermi-}LAT. At VHE energies, it is very challenging to characterize the diffuse emission component along the Galactic plane. However, the $\gamma$-ray spectrum of the nearby starburst galaxy NGC\,253 is best interpreted as the emission of the diffuse relativistic proton population in that galaxy, naturally with a CR density well beyond the Galactic one. An updated analysis of new H.E.S.S.\ and {\it Fermi-}LAT data nicely strengthens this hypothesis, details will be shown in an upcoming journal publication.

\section{Probing pulsar winds on different scales}

PWNe constitute the largest fraction of identified Galactic VHE sources, and are most likely the most numerous amongst the yet unidentified Galactic VHE sources as well. The VHE emission is commonly attributed to IC emission, PWNe are driven by relativistic electrons. While not contributing to the bulk of (hadronic) CRs, PWNe are nevertheless very important objects to study relativistic particle acceleration and transport. Most young and powerful pulsars are associated with detected PWNe~\cite{bib:hess2017pwnpop}.

\subsection{The VHE size of the Crab nebula}

The Crab nebula is well resolved and morphologically studied in most lower energy wavebands. The expected extension of the nebula in $\gamma$-rays is just at the edge of being detectable with VHE instruments. Using a very detailed (observation run-wise) simulation of the point spread function (psf) of the instrument, it could now be shown with H.E.S.S.\ data that sub-psf extensions can reliably be resolved, and that an intrinsic VHE source width of $\sigma \simeq 52''$ fits the Crab data best, with the measured $\sigma$ being significantly incompatible with zero~\cite{bib:hess2017crabextension}.

\subsection{Binary systems and pulsar winds}

A unique opportunity to probe pulsar winds is given when the objects are in binary systems. PSR\,B1259-63 is one of two $\gamma$-ray binary systems with a confirmed pulsar as compact object. The other $\gamma$-ray binaries may also host neutron stars, but confirmation is pending. The acceleration process in PSR\,B1259-63 is governed by the pulsar wind interacting with the stellar disk, while photon emission is affected by the stellar photon density (both as IC target and for $\gamma-\gamma$ absorption). H.E.S.S.\ observations around the 2014 periastron passage of PSR\,B1259-63 augmented the already known VHE orbital light curve, by showing significant VHE emission at the ``GeV flare'' period at 40-50 days after periastron, as well as by showing an unexpected high emission level also before the first disk crossing~\cite{bib:hess2017psrb1259}.

The number of known $\gamma$-ray binaries is very small. The latest H.E.S.S.\ highlight in that regard is the detection of VHE emission from the new binary system LMC\,P3 at binary phase $\sim$$0.3$~\cite{bib:hess2017lmcp3}, following the discovery of the system with {\it Fermi-}LAT. The light curve behaviour across different wavebands shares some similarities with LS\,5039. Despite being the most luminous $\gamma$-ray binary known to date, the VHE data from LMC\,P3 are however not sufficient to determine the binary period independently. This is in stark contrast to LS\,5039, where the period derived from H.E.S.S.\ data is the most accurate across all wavebands to date. The nature of the compact object in both of these systems is however not known yet.

\subsection{Extended VHE pulsar wind nebulae}

On larger angular scales, VHE-emitting PWNe can be used to explore the accelerated particle population, the mode of particle transport (e.g.\ diffusion vs.\ advection), and environmental conditions such as magnetic field turbulence away from the pulsar. For some objects, this can be done in great detail due to high VHE photon statistics using spectro-morphology. Three prominent sources, HESS\,J1825$-$137, MSH\,15$-$5{\it 2}, and Vela\,X, are currently in the focus of such studies~\cite{bib:hess2017j1825,bib:hess2017msh1552,bib:hess2017velax}.

\subsection{The puzzle of the lack of H.E.S.S.\ counterparts to some Galactic HAWC sources}

A comparison of the VHE emission from the Galactic plane measured with HAWC with that by H.E.S.S.\ has revealed some puzzling differences. Seven HAWC sources in the field of view of the HGPS have not been detected with H.E.S.S. Besides the possibility of attributing this discrepancy to different thresholds of the two data sets, a likely interpretation (at least for some of the sources) is that the HAWC data reveal largely extended (Geminga-like) PWNe, which are undetectable with H.E.S.S.\ due to their potentially low VHE surface brightness (e.g.~\cite{bib:hess2017j1928}).

\subsection{The local cosmic ray electron spectrum}

Due to strong energy losses in the ISM, the CR electron spectrum measured at Earth is a probe of the local surroundings of CR electron accelerators, with the size of the horizon inversely scaling with particle energy. With H.E.S.S., the electron spectrum could now be measured up to $\sim$$20$\,TeV. Above the known break at $\sim$$1$\,TeV, the spectrum appears feature-less. The spectrum can be used to constrain expectations from local accelerators such as nearby pulsars (i.e.\ Geminga-like PWNe) as well as SNRs. The spectrum is an important element in the discussion regarding the interpretation of the positron excess seen with PAMELA and AMS.

\section{Active Galactic Nuclei}

\subsection{The H.E.S.S.\ extragalactic scan}

The size of the FoV of H.E.S.S.\ does not permit surveys of large fractions of the extragalactic sky. Still, the combined fields of all extragalactic pointings performed so far with H.E.S.S.\ comprises coverage of a substantial fraction ($\sim$$6.5\%$) of the sky. A uniform reanalysis of the H.E.S.S.\ I extragalactic data (with $\sim$$2700$ hours of observations) constitutes the H.E.S.S.\ extragalactic scan (HEGS) and is used for population studies, transient searches, and variability studies. 216 3FGL sources have been serendipitously observed. For 48 of those sources, the VHE upper limits constrain a simple {\it Fermi-}LAT extrapolation (poster shown at this conference). 

\subsection{A measurement of the extragalactic background light}

Individual AGN VHE spectra can be used to constrain the density of EBL photons, by making assumptions about the intrinsic (unabsorbed) source spectra. Usually, the range of possible scalings of models for the EBL spectral density (as function of redshift) are constrained in that way. With a larger number of AGN source spectra, it was now possible to directly measure the spectral EBL density in a few wavelength ranges (i.e.\ without EBL model scaling), only assuming that the intrinsic VHE spectra are described by smooth concave shapes~\cite{bib:hess2017eblmeasurement}.

\subsection{Constraints on Inverse Compton peak positions}

Once the EBL level is understood to sufficient accuracy, the combination of {\it Fermi-}LAT and (deabsorbed) H.E.S.S.\ spectra can be exploited to derive constraints on the position of the IC peak positions of selected high-peaked BLLacs. Some spectra permit to measure the peak energy, while for some sources only lower limits can be derived~\cite{bib:hess2017intrinsicspectra}.

\subsection{Beyond BLLacs}

Due to the low energy threshold of CT5, low-frequency peaked AGN are a prime target for H.E.S.S.\ II observations. After ApLib, PKS\,1749+096 is the second low-peaked BLLac object from which VHE emission has been detected with H.E.S.S.\ now~\cite{bib:hess2017blazartoo}. Flat spectrum radio quasars (FSRQs) are particularly interesting (and challenging) because of the increased complexity when interpreting their spectra. Besides the ubiqutous EBL absorption, their VHE emission can also suffer from absorption due to strong emission from broad-line regions. In~\cite{bib:hess2017pks0736,bib:hess20173c279}, the spectra of PKS\,0736+017 and 3C\,279 have been examined also with the goal to put constraints on the location of the emitting region within the jets. Another FSRQ of interest is PKS\,1510-089, where observations have e.g.\ shown a (near-)orphan flare, i.e.\ a VHE flare with barely any counterpart in lower frequencies~\cite{bib:hess20171510flare}. 

Regarding radio galaxies, an updated combined {\it Fermi-}LAT -- H.E.S.S.\ spectrum of Cen\,A is under way, clearly confirming the connection of the H.E.S.S.\ spectrum with the second component emerging towards the high energy end of the {\it Fermi-}LAT spectrum. VHE emission from the radio galaxy PKS\,0625-354 has recently been discovered with H.E.S.S., the emission may however be attributed to a blazar core. 

The VHE data of selected (in particular hard-spectrum) sources can also be exploited to probe the density of the surrounding intergalactic magnetic field. A study of a potentially interesting source in that regard, 1RXS\,J023832.6-311658, has recently been conducted~\cite{bib:hess2017j023832}, showing however that the expected constraints from this source data are currently modest.

\subsection{Flares and variability}

Two strongly variable AGN have been at the focus of variability studies in the recent time, Mrk\,501 and 3C\,279. The flaring episode of Mrk\,501 already dates back to 2014, H.E.S.S.\ observations were triggered by an alert from the FACT telescope. Given the large zenith angles for H.E.S.S.\ observations for this Northern source, the energy threshold (even with H.E.S.S.\ II) is around 1\,TeV ($\sim$$64^{\circ}$ zenith angle). The H.E.S.S.\ flux points therefore match in energy band with the long-term monitoring light curve derived with FACT, but of course with much larger photon statistics. Preliminary results have been presented in~\cite{bib:hess2017mkn501}. 

3C\,279 (a highly variable FSRQ at redshift $z=0.54$) has shown in 2015 an extremely bright maximum with an increase of its flux level by a factor of $\sim$$10$ above 100\,MeV and a variability timescale of $\sim$$2$ minutes as detected with {\it Fermi-}LAT. The activity triggered H.E.S.S.\ observations, the H.E.S.S.\ II mono spectrum (with a threshold of 66\,GeV) derived from $\sim$$3$ hours of observations in one night where the source was detected with H.E.S.S.\ reveals a very steep VHE spectrum (photon index larger than 4), strongly affected by EBL absorption and potentially also by internal absorption~\cite{bib:hess20173c279}. 

Exploiting the strong VHE variability, both sources have -- amongst other studies -- also been used to derive limits on a possible violation of Lorentz invariance, following similar studies of other sources during the past years.

\section{Indirect dark matter searches}

\begin{figure}[t]
    \centering
	\includegraphics[width=0.46\textwidth]{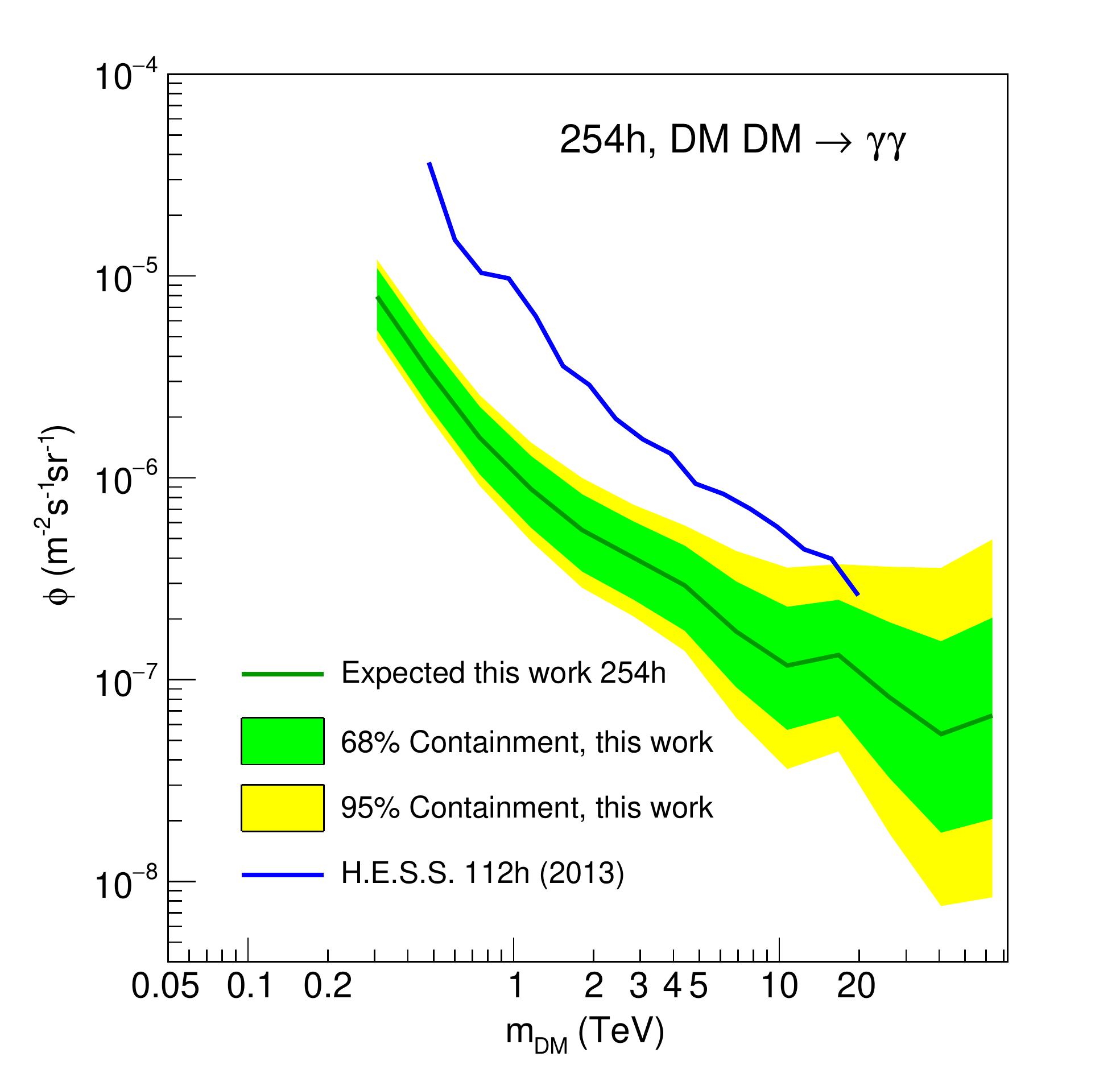}
    \qquad
	\includegraphics[width=0.46\textwidth]{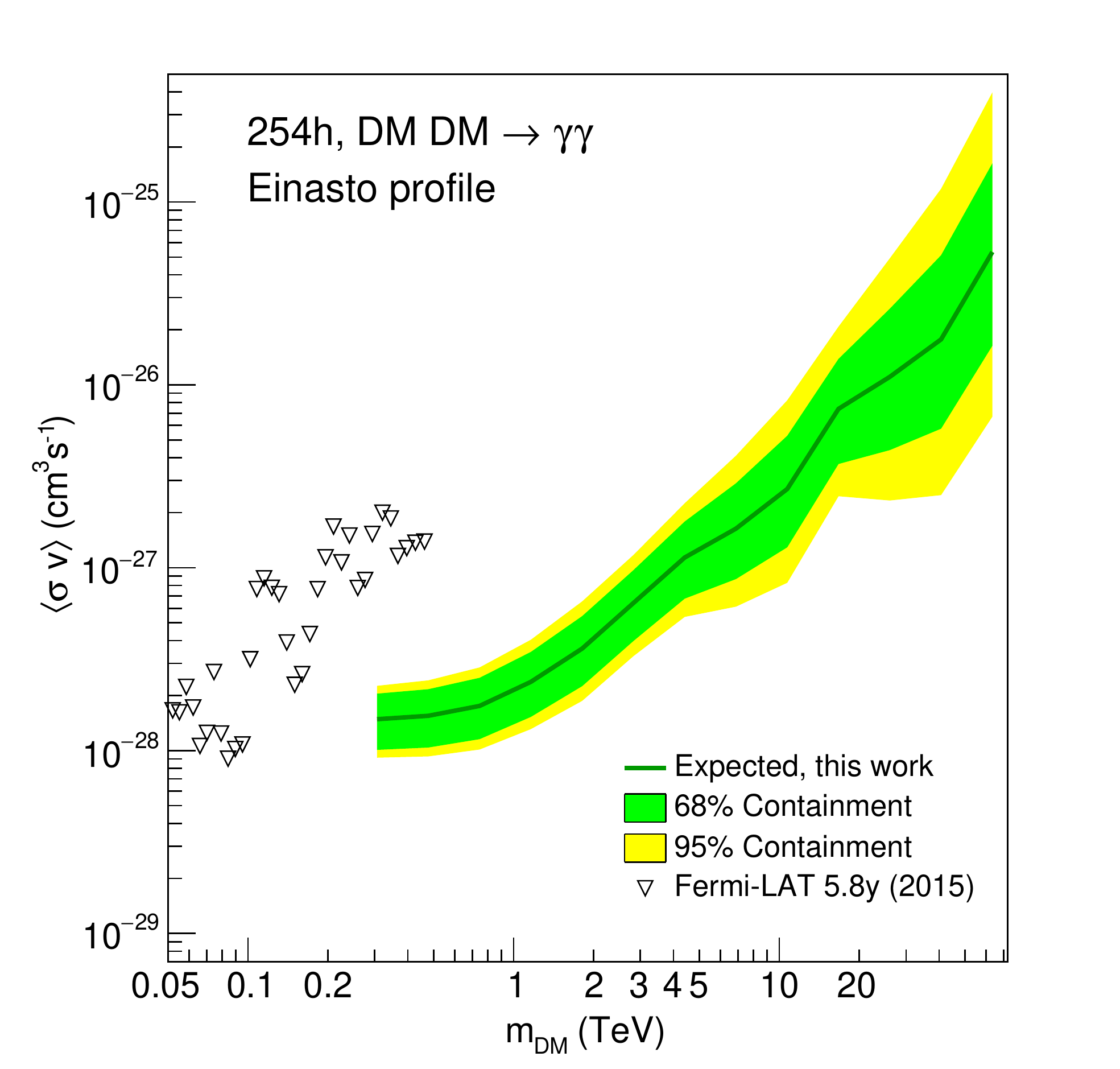}
	\caption{
New constraints on the dark matter annihilation into two photons ($\mathrm{DM~DM} \rightarrow \gamma\gamma$) as a function of the dark matter mass $m_\mathrm{DM}$, derived from observations of the Galactic center halo with H.E.S.S. Shown are mean expected limits (green solid line) with their 68\% (light green box) and 95\% (yellow box) containment bands. Previous H.E.S.S.\ limits published in 2013 are also indicated as blue solid line. {\it Left:} 95\% c.l.\ upper limits on the photons flux $\mathrm{\phi}$ from dark matter annihilations. {\it Right:} 95\% c.l.\ upper limits on the thermally-averaged velocity-weighted annihilation cross section $\langle\sigma v\rangle$ for the Einasto profile, compared to {\it Fermi}-LAT limits. Figure taken from \cite{bib:hess2017halodmlimits}.
	}
	\label{Fig:HESSGChalo}
\end{figure}

Dark matter particles may reveal themselves through self-annihilation and subsequent $\gamma$-ray emission. Current prime targets for the search for emission from this process with H.E.S.S.\ comprise the Galactic center halo~\cite{bib:hess2017halodmlimits} and selected dwarf spheroidals~\cite{bib:hess2017dwarfdmlimits}. Deep observations and optimized statistical strategies are key ingredients to the program. The annihilation process may lead to line-like features in the spectra. New upper limits for line-like emission have been obtained from these sources, for the Galactic halo region nicely complementing the limits derived with {\it Fermi-}LAT towards lower masses of the hypothetical dark matter particles (\cite{bib:hess2017halodmlimits}, see Fig.\,\ref{Fig:HESSGChalo}).

\section{Discussion}

In the discussion after the presentation, the question was raised what limits the extension of the H.E.S.S.\ II spectrum towards lower energies, into the {\it Fermi-}LAT range. In fact, increasing the H.E.S.S.\ II exposure time does at some stage not improve the spectrum at the lowest energies, since systematic errors will start to dominate. Another question concerned the validity of the spectral extrapolation from the {\it Fermi-}LAT into the H.E.S.S.\ energy band for the HEGS sources, since the observations are not contemporaneous. This is indeed an obvious shortcoming, which can to some extent be overcome by looking at the properties of a sample of sources instead of individual sources, assuming that the exposures of the individual sources were taken in an unbiased way.

\section*{Acknowledgements}

{\small 
The support of the Namibian authorities and of the University of Namibia in facilitating the construction and operation of H.E.S.S.\ is gratefully acknowledged, as is the support by the German Ministry for Education and Research (BMBF), the Max Planck Society, the German Research Foundation (DFG), the Alexander von Humboldt Foundation, the Deutsche Forschungsgemeinschaft, the French Ministry for Research, the CNRS-IN2P3 and the Astroparticle Interdisciplinary Programme of the CNRS, the U.K.\ Science and Technology Facilities Council (STFC), the IPNP of the Charles University, the Czech Science Foundation, the Polish National Science Centre, the South African Department of Science and Technology and National Research Foundation, the University of Namibia, the National Commission on Research, Science \& Technology of Namibia (NCRST), the Innsbruck University, the Austrian Science Fund (FWF), and the Austrian Federal Ministry for Science, Research and Economy, the University of Adelaide and the Australian Research Council, the Japan Society for the Promotion of Science and by the University of Amsterdam. We appreciate the excellent work of the technical support staff in Berlin, Durham, Hamburg, Heidelberg, Palaiseau, Paris, Saclay, and in Namibia in the construction and operation of the equipment. This work benefited from services provided by the H.E.S.S.\ Virtual Organisation, supported by the national resource providers of the EGI Federation.
}

\end{document}